\documentclass[aps,prb,notitlepage,superscriptaddress,twocolumn]{revtex4-2}

\usepackage{amsmath}
\usepackage{amssymb}
\usepackage{graphicx}
\usepackage{epstopdf}
\usepackage[colorlinks=true]{hyperref}

\newcommand{\sech}{\mathrm{sech}}

\begin{document}
\title{The dynamics of a domain wall in ferrimagnets driven by spin-transfer torque}

\author{Dong-Hyun Kim}
\affiliation{Department of Semiconductor Systems Engineering, Korea University, Seoul 02841, Republic of Korea}
\author{Duck-Ho Kim}
\affiliation{Center for Spintronics, Korea Institute of Science and Technology, Seoul 136-791, Republic of Korea}
\author{Kab-Jin Kim}
\affiliation{Department of Physics, KAIST, Daejeon 34141, Republic of Korea}
\author{Kyoung-Woong Moon}
\affiliation{Quantum Technology Institute, Korea Research Institute of Standards and Science, Daejeon 34113, Republic of Korea}
\author{Seungmo Yang}
\affiliation{Quantum Technology Institute, Korea Research Institute of Standards and Science, Daejeon 34113, Republic of Korea}
\author{Kyung-Jin Lee}
\affiliation{Department of Semiconductor Systems Engineering, Korea University, Seoul 02841, Republic of Korea}
\affiliation{Department of Materials Science and Engineering, Korea University, Seoul 02841, Republic of Korea}
\affiliation{KU-KIST Graduate School of Converging Science and Technology, Korea University, Seoul 02841, Republic of Korea}
\author{Se Kwon Kim}
\affiliation{Department of Physics and Astronomy, University of Missouri, Columbia, Missouri 65211, USA}

\date{\today}

\begin{abstract}
The spin-transfer-torque-driven (STT-driven) dynamics of a domain wall in an easy-axis rare-earth transition-metal ferrimagnet is investigated theoretically and numerically in the vicinity of the angular momentum compensation point $T_A$, where the net spin density vanishes. The particular focus is given on the unusual interaction of the antiferromagnetic dynamics of a ferrimagnetic domain wall and the adiabatic component of STT, which is absent in antiferromagnets but exists in the ferrimagnets due to the dominant coupling of conduction electrons to transition-metal spins. Specifically, we first show that the STT-induced domain-wall velocity changes its sign across $T_A$ due to the sign change of the net spin density, giving rise to a phenomenon unique to ferrimagnets that can be used to characterize $T_A$ electrically. It is also shown that the frequency of the STT-induced domain-wall precession exhibits its maximum at $T_A$ and it can approach the spin-wave gap at sufficiently high currents. Lastly, we report a numerical observation that, as the current density increases, the domain-wall velocity starts to deviate from the linear-response result, calling for a more comprehensive theory for the domain-wall dynamics in ferrimagnets driven by a strong current.
\end{abstract}

\maketitle

\section{Introduction}

Spintronics is the field, in which electrons' spin is utilized in addition to charge for the advancement of information processing technology beyond the conventional charge-based electronics, and, therefore, the interaction between charge and spin has been one of the central topics in the field. In particular, the effect of a charge current on the magnetic dynamics, which is described as the spin-transfer torque (STT), has been intensively studied for metallic ferromagnets since the first theoretical predictions in 1996~\cite{SlonczewskiJMMM1996, BergerPRB1996}. One of the major practical utilities of STT in spintronics is to drive a magnetic domain wall, a topological defect between two uniform domains, that can be used to realize racetrack memory~\cite{ParkinScience2008}. Also, fundamental research on STT-induced domain-wall motion has been allowing us to strengthen our understanding of spin-charge interaction in ferromagnets~\cite{LiPRL2004, ThiavilleEPL2005, TataraPR2008, BurrowesNP2009}.

Departing from ferromagnets consisting of parallel spins, there is another class of magnets called antiferromagnets, where neighboring spins are antiparallel. Antiferromagnets have been attracting great attention in spintronics, particularly for the last decade, owing to their much faster dynamics than ferromagnets and the ensuing promise for ultrafast spintronic devices~\cite{JungwirthNN2016}. However, the research on antiferromagnetic dynamics has been impeded by difficulties in controlling and detecting them due to, e.g., the absence of magnetization. For this reason, the previous research on STT in antiferromagnets has been mostly theoretical~\cite{IvanovLTP2005, NunezPRB2006, HalsPRL2011, VelkovNJP2016, ZeleznyNP2018, GalkinaLTP2018}.

There have been recent developments in understanding and utilizing antiferromagnetic dynamics enabled by an emerging class of magnets called ferrimagnets, which consist of two or more inequivalent magnetic sublattices coupled antiferromagnetically~\cite{KirilyukRMP2010}. These ferrimagnets, which are typified by rare-earth transition-metal (RE-TM) ferrimagnetic alloys such as GdCo or GdFeCo, can exhibit antiferromagnetic dynamics owing to the antiferromagnetic coupling of constituent spins, and, at the same time, can be controlled and detected easily due to small, but finite magnetization caused by imperfect cancellation of neighboring magnetic moments. This unique combination of antiferromagnetic dynamics and finite magnetization has recently allowed for achieving fast domain-wall motion~\cite{KimNM2017, CarettaNN2018, SiddiquiPRL2018, CaiNE2020} and magnetization switching~\cite{OstlerNC2012, FinleyPRA2016, MishraPRL2017} in various ferrimagnets. In particular, \textcite{OkunoNE2019} recently reported an experimental study of STT in ferrimagnets through current-assisted field-driven domain-wall motion, introducing ferrimagnets as a useful platform to investigate STT in magnets with antiferromagnetic coupling~\cite{LavrijsenNE2019}. 

In this work, we theoretically and numerically study the domain-wall dynamics in RE-TM ferrimagnets driven by STT, which has not been explored yet. One of the unique features of ferrimagnets that are absent in ferromagnets and antiferromagnets is that their spin density can be continuously tuned across zero by changing temperature or chemical composition. The temperature at which the net spin density vanishes is called the angular momentum compensation point $T_A$~\cite{StanciuPRB2006, BinderPRB2006}, and it offers one of the situations where the advantage of ferrimagnets is most prominent: their dynamics is purely antiferromagnetic due to the vanishing spin density and thus is shown to be the fastest~\cite{KimNM2017, CarettaNN2018, SiddiquiPRL2018}. Therefore, in our study on STT-driven domain-wall dynamics, we will focus on ferrimagnets in the vicinity of the angular momentum compensation point $T_A$.

\begin{figure}
\includegraphics[width=\columnwidth]{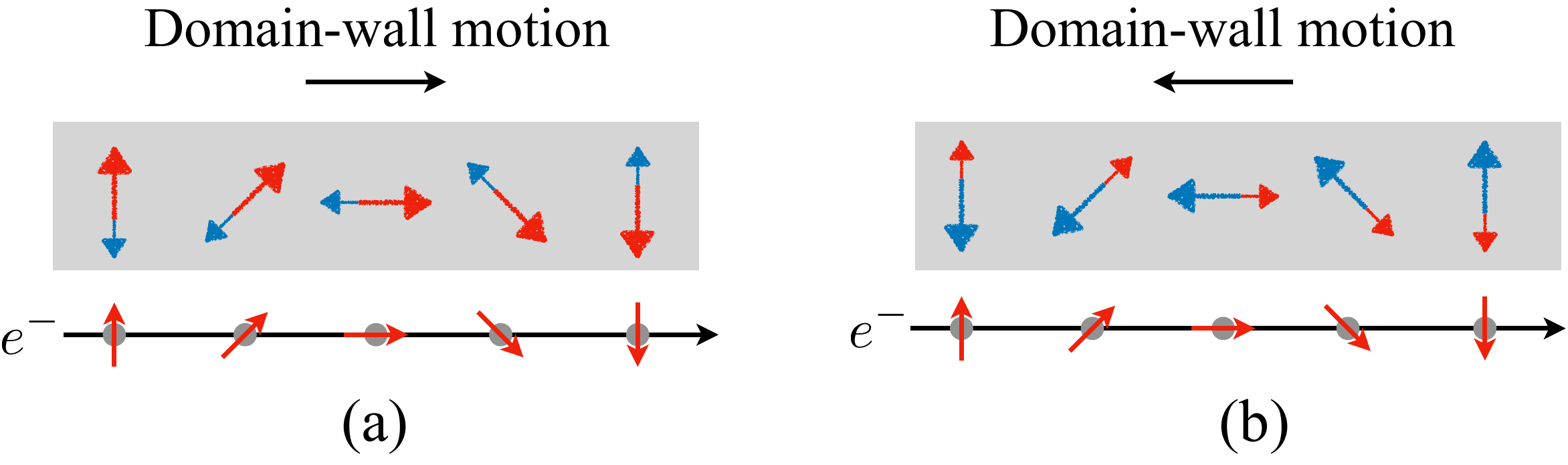}
\caption{Schematic illustration of a domain-wall motion driven by the adiabatic STT in a RE-TM ferrimagnet. The red and the blue arrows in the gray box represent spins of TM and RE elements, respectively. When a conduction electron (denoted by $e^{-}$) traverses the domain wall from the left to the right, its spin denoted by the red arrow follows the TM spin direction adiabatically. After passing through the domain wall, the change of electron's spin is transferred to the domain wall via the adiabatic STT. (a) When TM spins are larger than RE spins (corresponding to $T > T_A$), the net spin direction is given by the TM spin direction. The transfer of up-spin from conduction electrons to the spin texture expands the left domain with net-spin up, pushing the domain wall to the right. (b) When RE spins are larger than TM spins (corresponding to $T < T_A$), the net spin direction is given by the RE spin direction. The transfer of up-spin from conduction electrons drive the domain wall to the left by expanding the right domain with net-spin up.}
\label{fig:fig1}
\end{figure}

STT of ferrimagnets that describes the effect of a current on a spatially varying spin texture is similar to STT of ferromagnets due to the dominant coupling of conduction electrons' spin to one of multiple sublattices of ferrimagnets, as have been invoked in Ref.~\cite{KimPRB2017}, where the STT-driven dynamics are studied for two-dimensional spin textures called skyrmions in ferrimagnets. It consists of the reactive and the dissipative components, which are also referred to as the adiabatic and the nonadiabatic STT. The adiabatic STT, which is even under time reversal, describes the angular momentum transfer from conduction electrons to the background spin texture such as a domain wall when electrons' spin follows the local spin texture adiabatically. The nonadiabatic STT captures the angular momentum transfer via the other processes deviating from the aforementioned adiabatic process, e.g., mistracking between conduction electrons' spin and the spin texture. While both terms exist in ferromagnets, antiferromagnets are lack of the adiabatic STT since electrons' spin cannot follow atomically-changing staggered spins adiabatically~\cite{HalsPRL2011, TvetenPRL2013}. STT of RE-TM ferrimagnets possesses both terms akin to ferromagnets since conduction electrons are known to interact mostly with TM magnetic moments and thus are spin-polarized following TM's magnetization~\cite{MartinPRB2008, OkunoAPE2016}. See Fig.~\ref{fig:fig1} for the illustration. Therefore, in the vicinity of angular momentum compensation point, where the nature of dynamics is antiferromagnetic but STT possesses the adiabatic term, ferrimagnets are expected to exhibit a unique phenomena that can occur neither in ferromagnets nor in antiferromagnets, which we reveal theoretically and numerically in this work through the domain-wall dynamics.

This paper is organized as follows. In Sec.~\ref{sec:sec2}, we study the STT-driven domain-wall dynamics in ferrimagnets by varying the temperature across the angular momentum compensation point $T_A$. Specifically, by studying the dependence of the velocity and the angular velocity of a domain wall on the net spin density and the current within the linear response, we show that the domain-wall velocity changes its sign across $T_A$ (see Fig.~\ref{fig:fig1} for the illustration), offering an electrical way to identify $T_A$ that is known to be difficult to characterize. In addition, the angular velocity is shown to exhibit its maximum at $T_A$, where most of the transferred angular momentum from conduction electrons is used for rotating the domain wall by accumulating a nonequilibrium spin density inside it. The theoretical result based on the collective coordinate approach is supported by atomistic spin simulations. In Sec.~\ref{sec:sec3}, we numerically study the STT-driven dynamics of a domain wall exactly at $T_A$ by applying large currents to go beyond the linear-response regime. As the current increases, the domain-wall angular velocity is shown to saturate to the spin-wave gap, which is caused by the increase of the domain-wall width. In addition, at large currents, we observe that the domain-wall velocity deviates from what is predicted from the linear-response theory, showing a limitation of the linear-response theory for the domain-wall dynamics at high biases. We conclude the paper by providing a summary and future outlook in Sec.~\ref{sec:sec4}.

\section{STT-driven domain-wall dynamics within the linear response}
\label{sec:sec2}

In this section, we study the dynamics of a domain wall in ferrimagnets driven by STT within the linear response, i.e., at sufficiently small currents. For concreteness, we consider RE-TM ferrimagnets which are composed of two antiferromagnetically-coupled sublattices of RE spins and TM spins.

\subsection{Theory}

The Landau-Lifshitz-Gilbert-like equation for a RE-TM ferrimagnet with STT is given by~\cite{IvanovJETP1983, BaryakhtarSPU1985, ChioleroPRB1997, KimPRB2017, TserkovnyakJMMM2008, OkunoNE2019, IvanovLTP2019}
\begin{equation}
\begin{split}
& \delta_s \dot{\mathbf{n}} - \alpha s \mathbf{n} \times \dot{\mathbf{n}} - \rho \mathbf{n} \times \ddot{\mathbf{n}} \\
& = - \mathbf{n} \times \mathbf{h}_\text{eff} + P (\mathbf{J} \cdot \boldsymbol{\nabla}) \mathbf{n} - \beta P \mathbf{n} \times (\mathbf{J} \cdot \boldsymbol{\nabla}) \mathbf{n} \, , 
\end{split}
\label{eq:llg}
\end{equation}
to linear order in the bias, a charge current density $\mathbf{J} = J \hat{\mathbf{x}}$, where $\mathbf{n}$ is the unit vector along the magnetization direction of RE elements, $\delta_s$ is the equilibrium spin density along $- \mathbf{n}$ (i.e., along the spin direction of RE elements), $\alpha > 0$ is the Gilbert damping constant, $s$ is the sum of the spin densities of the two sublattices, $\rho$ is the moment of inertia representing antiferromagnetic dynamics of $\mathbf{n}$~\cite{BaltzRMP2018}, $\mathbf{h}_\text{eff} \equiv - \delta U / \delta \mathbf{n}$ is the effective field conjugate to $\mathbf{n}$, and $U[\mathbf{n}]$ is the potential energy. The last two terms on the right-hand side are the adiabatic and the nonadiabatic STT terms, where $P$ is the spin conversion factor given by $P = (\hbar / 2 e) (\sigma_\uparrow - \sigma_\downarrow) / (\sigma_\uparrow + \sigma_\downarrow)$ (with the electron charge $e > 0$) which characterizes the polarization of the spin-dependent conductivity $\sigma_s$ ($s = \uparrow$ or $\downarrow$ with $\uparrow$ chosen along $-\mathbf{n}$), and $\beta$ is the dimensionless parameter characterizing the nonadiabatic torque term. Note that there exists the adiabatic component of STT since the charge current can be spin-polarized according to one of two sublattices, which is in contrast with STT for antiferromagnets where the adiabatic component is absent~\cite{HalsPRL2011, TvetenPRL2013}. 

We consider a quasi-one-dimensional magnet with uniaxial anisotropy described by the following potential energy:
\begin{equation}
\label{eq:U}
U = \int dV (A \mathbf{n}'^2 - K m_z^2) / 2 \, ,
\end{equation}
which has been used to describe the domain-wall motion in magnets with perpendicular magnetic anisotropy and negligible in-plane anisotropy (see, e.g., Refs.~\cite{TvetenPRL2014, KimPRB2014, CarettaNN2018, YangPRB2019}). Here, $A$ is the exchange coefficient, $K$ is the easy-axis anisotropy coefficient, and $'$ represents the derivative with respect to the spatial coordinate $x$. Here, we assume that the magnetic order is uniform in the $yz$ plane. A domain wall is a topological soliton connecting two ground states $\mathbf{n} = \pm \hat{\mathbf{z}}$. Its low-energy dynamics is known to be well described by two collective coordinates, position $X(t)$ and angle $\Phi(t)$, via the so-called Walker ansatz~\cite{SchryerJAP1974}: $\mathbf{n}(x, t) = \{\sech((x - X) / \lambda) \cos \Phi, \sech((x - X) / \lambda) \sin \Phi, - \tanh ((x - X) / \lambda)\}$, where $\lambda$ represents the domain-wall width. In equilibrium, the domain-wall width is given by $\lambda_0 = \sqrt{A/K}$ determined by the competition between the exchange energy and the anisotropy.

By employing the collective-coordinate approach~\cite{ThielePRL1973, ThiavilleEPL2005, TretiakovPRL2008}, we obtain the equations of motion for $X$ and $\Phi$, which are given by
\begin{eqnarray}
\delta_s \lambda \dot{\Phi} + \alpha s \dot{X} + \rho \ddot{X} &=& - \beta P J \, , \\
\delta_s \dot{X} - \alpha s \lambda \dot{\Phi} - \rho \lambda \ddot{\Phi} &=& - P J \, .
\end{eqnarray}
The steady-state velocity $V$ and the angular velocity $\Omega$ are then given by, respectively,
\begin{equation}
\label{eq:V}
\dot{X} \rightarrow V = - \frac{P J (\delta_s + \alpha \beta s)}{\delta_s^2 + (\alpha s)^2} \, ,
\end{equation}
and
\begin{equation}
\label{eq:Omega0}
\dot{\Phi} \rightarrow \Omega = \frac{P J}{\lambda_0} \frac{\alpha s - \beta \delta_s}{\delta_s^2 + (\alpha s)^2} \, .
\end{equation}
This is our main analytical result: The domain-wall velocity $V$ and the angular velocity $\Omega$ as a function of the net spin density $\delta_s$, which can be controlled in RE-TM ferrimagnets by changing temperature or chemical composition.

Let us discuss the obtained results for specific cases. First, when the net spin density is sufficiently large, i.e., when the temperature is sufficiently away from $T_A$, the domain-wall velocity can be approximated by
\begin{equation}
V \approx - P J / \delta_s \, , \quad \text{for } |\delta_s| \gg \alpha s \, \, ,
\end{equation}
while assuming $|\beta| \ll 1$. This can be understood as the result of the angular-momentum transfer $\propto P J$ from conduction electrons to the domain wall via the adiabatic STT. The absorption of the transferred angular momentum translates into the expansion of one of the two domains at the velocity $V$~\cite{SlonczewskiJMMM1996, BergerPRB1996}. Note that the direction of the domain-wall motion changes when the sign of the net spin density changes, i.e., across $T_A$. The net spin density $\delta_s$ is defined with respect to $- \mathbf{n}$, and thus, in our domain-wall ansatz, the net spin densities of the left ($x \rightarrow - \infty$) and the right ($x \rightarrow + \infty$) domains are given by $+ \delta_s$ and $- \delta_s$, respectively, polarized in the $+z$ direction. Therefore, for the given angular momentum transfer from conduction electrons, whether the left domain or the right domain expands is determined by the sign of the net spin density. See Fig.~\ref{fig:fig1} for the schematic illustration. We would like to remark here that the analogous result of the reversal of the domain-wall motion has been reported in the theoretical study of the spin-wave-driven ferrimagnetic domain-wall motion~\cite{OhPRB2019}.

Second, when the net spin density vanishes $\delta_s = 0$, i.e., at $T_A$, the domain-wall velocity is reduced to 
\begin{equation}
V = - \beta P J / \alpha s \, , \quad \text{for } \delta_s = 0 \, .
\end{equation}
This reproduces the known result for the STT-induced domain-wall motion in antiferromagnets~\cite{HalsPRL2011}, where the domain wall is driven by the nonadiabatic STT $\propto \beta$. Also, for $\delta_s = 0$, the angular velocity is reduced to 
\begin{equation}
\label{eq:Omegata}
\Omega = P J / \alpha \lambda_0 s \, , \quad \text{for } \delta_s = 0 \, .
\end{equation}
This can be understood as follows. Conduction electrons transfer angular momentum at the rate $\propto PJ$ to the domain wall by passing through it. When the net spin density is finite $\delta_s \neq 0$, the domain wall can absorb the transferred angular momentum by moving, i.e., by expanding one of the two domains. However, when the net spin density vanishes $\delta_s = 0$, the transferred angular momentum cannot be absorbed by the domain-wall motion and thus it is accumulated inside the domain wall. This nonequilibrium spin density exerts the effective magnetic field~\footnote{Microscopically, the nonequilibrium spin density and the resultant effective magnetic field are the manifestiations of relative canting of the spin densities of multiple sublattices. See Refs.~\cite{TakeiPRB2014, JungwirthNN2016, BaltzRMP2018} or Supplemental Material of Ref.~\cite{KimPRB2017} for the detailed discussions.}, which in turn rotates the magnetic order inside the domain wall at the angular velocity $\Omega \propto P J$. The steady-state amount of the nonequilibrium spin density and the corresponding precession frequency $\Omega$ are determined by balancing the spin dissipation rate caused by the precession $\propto \alpha s \Omega$ and the generation rate of the nonequilibrium spin density $\propto P J$. The similar phenomenon has been observed numerically and explained theoretically in the dynamics of a domain wall in an antiferromagnet driven by a circularly-polarized magnon current~\cite{TvetenPRL2014, KimPRB2014}.

\subsection{Simulation}

To confirm the obtained analytical results, we perform numerical simulations by solving the following coupled atomistic LLG equations for RE-TM ferrimagnets~\cite{KimNM2017, OhPRB2017, OhPRB2019}:
\begin{equation}
\begin{split}
\frac{\partial \mathbf{A}_k}{\partial t} = & - \gamma_\text{re} \mathbf{A}_k \times \mathbf{H}^k_\text{eff,A} + \alpha_\text{re} \mathbf{A}_k \times \frac{\partial \mathbf{A}_k}{\partial t} \\
& - b_\text{re} \frac{\partial \mathbf{A}_k}{\partial x} - \beta_\text{re} b_\text{re} \mathbf{A}_k \times \frac{\partial \mathbf{A}_k}{\partial x} \, , \\
\frac{\partial \mathbf{B}_k}{\partial t} = & - \gamma_\text{tm} \mathbf{B}_k \times \mathbf{H}^k_\text{eff,B} + \alpha_\text{tm} \mathbf{B}_k \times \frac{\partial \mathbf{B}_k}{\partial t} \\
& - b_\text{tm} \frac{\partial \mathbf{B}_k}{\partial x} - \beta_\text{tm} b_\text{tm} \mathbf{B}_k \times \frac{\partial \mathbf{B}_k}{\partial x} \, ,
\end{split}
\end{equation}
where $\mathbf{A}_k$ and $\mathbf{B}_k$ are the normalized spins at the $k$th sites in RE and TM sublattices, respectively, $\mathbf{H}^k_\text{eff,A} = (1/\mu_\text{re}) \cdot \partial H / \partial \mathbf{A}_k$ and $\mathbf{H}^k_\text{eff,B} = (1/\mu_\text{tm}) \cdot  \partial H / \partial \mathbf{B}_k$ are the effective magnetic fields, $\mu_\text{re}$ and $\mu_\text{tm}$ are the local magnetic moments, $\alpha_\text{re}$ and $\alpha_\text{tm}$ are the Gilbert damping constants, $\gamma_\text{re}$ and $\gamma_\text{tm}$ are the gyromagnetic ratios, $M_\text{re}$ and $M_\text{tm}$ are the saturation magnetizations, $b_\text{re} = - g_\text{re} \mu_B P_\text{re} J / (2 e M_\text{re})$ and $b_\text{tm} = - g_\text{tm} \mu_B P_\text{tm} J / (2 e M_\text{tm})$ are the STT parameters~\footnote{The STT parameters $b_\text{re}$ and $b_\text{tm}$ have units of speed.}, $J$ is the charge current density, $e$ is the electron charge, $g_\text{re}$ and $g_\text{tm}$ are the g-factors, $P_\text{re}$ and $P_\text{tm}$ are the dimensionless spin polarizations, and $\beta_\text{re}$ and $\beta_\text{tm}$ are the dimensionless nonadiabatic STT parameters. Here, $H$ is the discrete Hamiltonian given by $H = A_\text{sim} \sum_k (\mathbf{A}_ k \cdot \mathbf{B}_k + \mathbf{B}_{k} \cdot \mathbf{A}_{k+1}) - K_\text{sim} \sum_k [(\mathbf{A}_k \cdot \hat{\mathbf{z}})^2 + (\mathbf{B}_k \cdot \hat{\mathbf{z}})^2]$.

\begin{figure}
\includegraphics[width=\columnwidth]{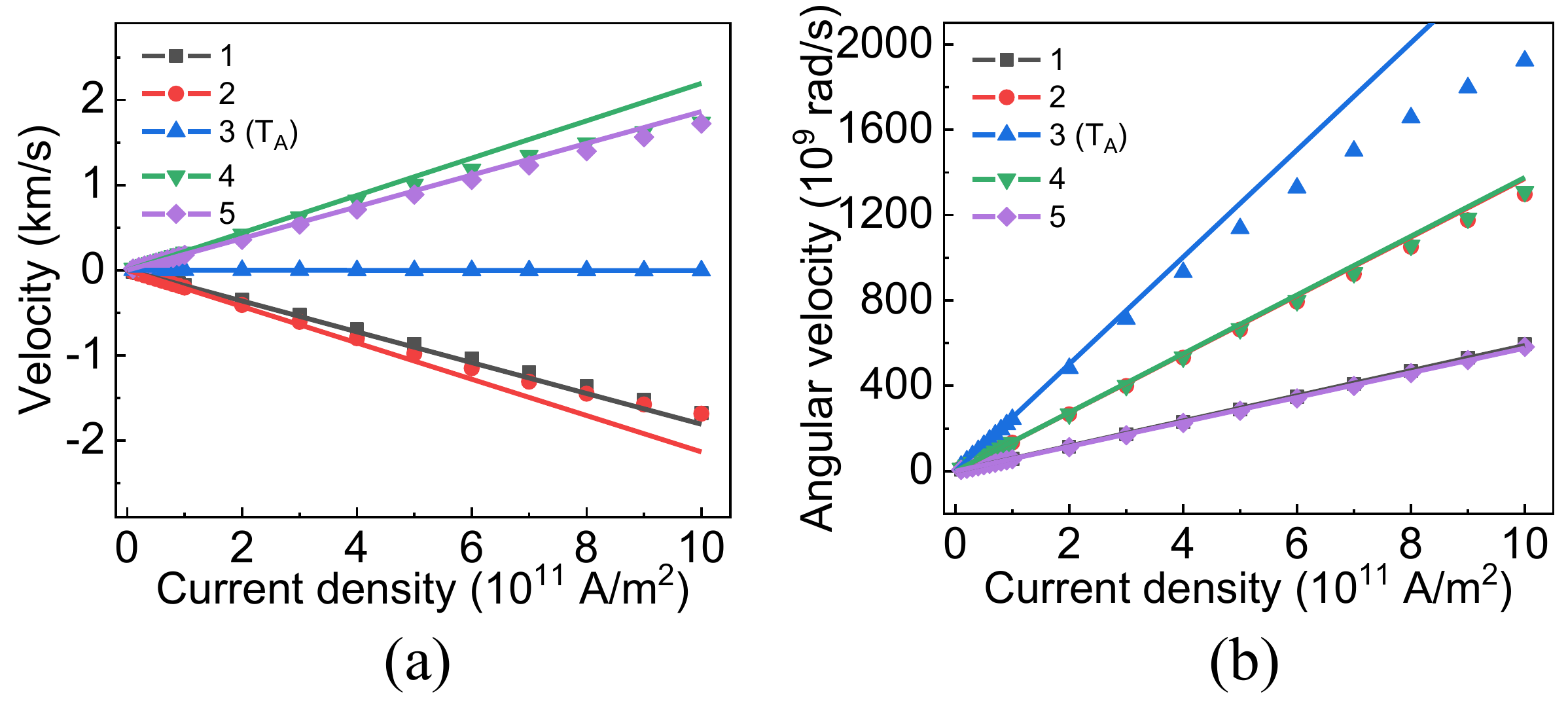}
\caption{(a) The domain-wall velocity and (b) the domain-wall angular velocity as a function of the current density $J \le 10^{12}$ A/m$^2$ at various temperatures shown in Table~\ref{tab:tab1}. The symbols are numerical results. The lines show the analytical results for the velocity $V$ [Eq.~(\ref{eq:V})] and the angular velocity $\Omega$ [Eq.~(\ref{eq:Omega0})] obtained within the linear response.}
\label{fig:fig2}
\end{figure}

\begin{table}[b]
\centering
\begin{tabular}{c | c | c | c | c | c}
\hline \hline
Index & 1 & 2 & 3 ($T_A$) & 4 & 5 \\
\hline 
$M_\text{re}$ (kA/m)	& 1020 & 1010 & 1000 & 990 & 980 \\
$M_\text{tm}$ (kA/m)	& 1130 & 1115 & 1100 & 1085 & 1070 \\
$\delta_s$ ($10^{-8}$J$\cdot$s/m$^3$) & -4.13 & -2.07 & 0 & 2.07 & 4.13 \\
\hline \hline
\end{tabular}
\caption{The values of the magnetic moments $M_\text{re}$ and $M_\text{tm}$ for transition-metal and rare-earth elements, respectively, and the net spin density $\delta_s$ used in atomistic spin simulations. Index 3 represents the angular momentum compensation point $T_A$.}
\label{tab:tab1}
\end{table}

For the sample geometry, we considered $3200 \times 100 \times 0.4$ nm$^3$ with cell size $0.4 \times 100 \times 0.4$ nm$^3$. Correspondingly, the lattice constant in the $x$ direction is given by $d = 0.4$ nm. The used material parameters are $A_\text{sim} = 7.5$ meV, $K_\text{sim} = 0.4$ meV, $\alpha_\text{tm} = \alpha_\text{re} = 0.002$, and the gyromagnetic ratios $\gamma_\text{re} = 1.76 \times 10^{11}$ \text{rad}/s$\cdot$T and $\gamma_\text{tm} = 1.936 \times 10^{11}$ \text{rad}/s$\cdot$T (corresponding to the g-factors $g_\text{tm} = 2.2$ and $g_\text{re} = 2$~\cite{BinderPRB2006}). The used magnetic moments for five different cases are shown in Table~\ref{tab:tab1}. For STT parameters, a current in RE-TM ferrimagnets is known to interact mostly with TM magnetic moments~\cite{MartinPRB2008, OkunoAPE2016}. Therefore, in this work, we consider the simplest case where the current interacts only with TM elements. Accordingly, we use the following STT parameters: $P_\text{re} = 0, P_\text{tm} = 0.3$, and $\beta_\text{tm} = 0.001$. Corresponding parameters in the continuum model [Eq.~(\ref{eq:U})] are given by $A = 4 A_\text{sim}/d$ and $K = 4 K_\text{sim}/d^3$. In addition, we have the following relations: $\mathbf{n} = (\mathbf{A}_k - \mathbf{B}_k) / |\mathbf{A}_k - \mathbf{B}_k|$, $s_\text{re} = M_\text{re} / \gamma_\text{re}$, $s_\text{tm} = M_\text{tm} / \gamma_\text{tm}$, $\delta_s = s_\text{re} - s_\text{tm}$, $s = s_\text{re} + s_\text{tm}$, $\alpha = (\alpha_\text{re} s_\text{re} + \alpha_\text{tm} s_\text{tm}) / s$, $P J = s (b_\text{re} - b_\text{tm})/2$, and $\beta = (\beta_\text{re} b_\text{re} + \beta_\text{tm} b_\text{tm}) / 2PJ$.

\begin{figure}
\includegraphics[width=\columnwidth]{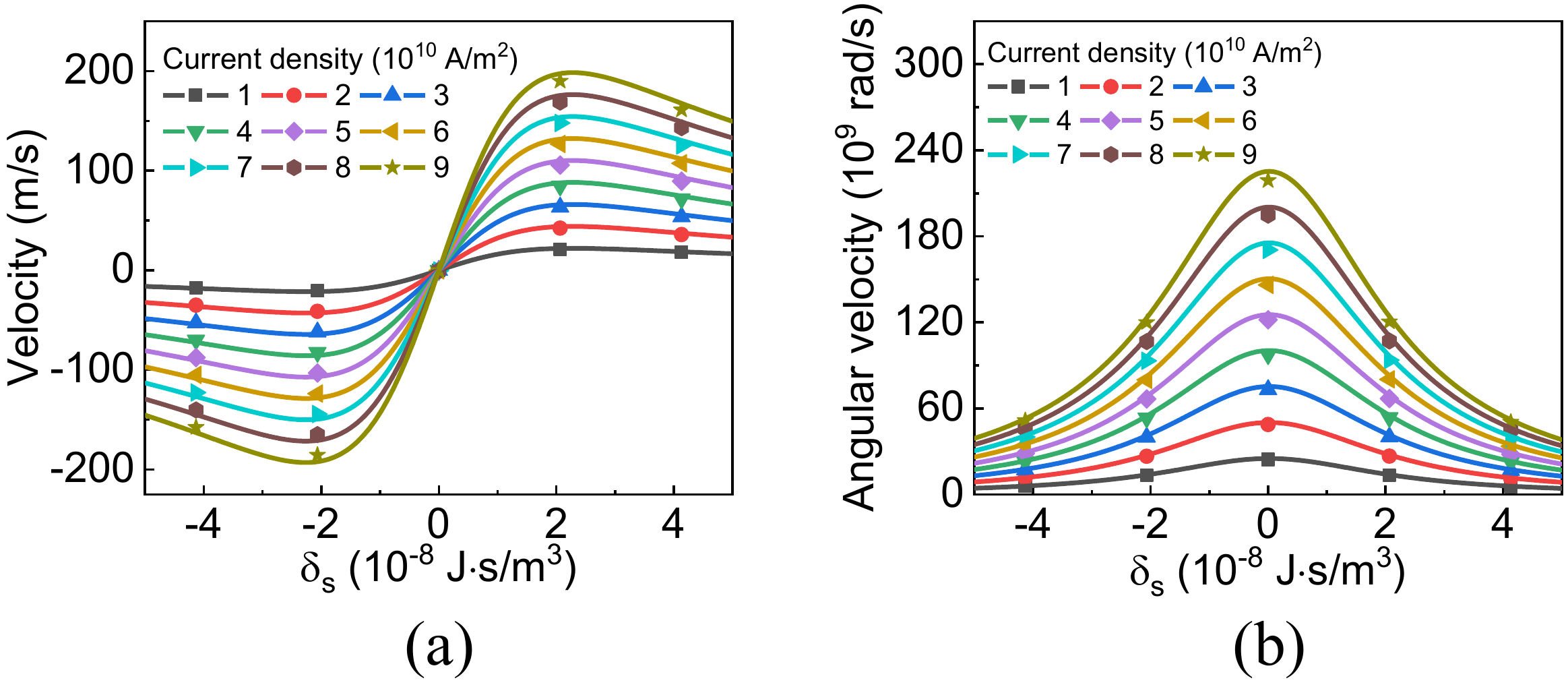}
\caption{(a) The domain-wall velocity and (b) the domain-wall angular velocity as functions of the spin density $\delta_s$ at various current densities. The symbols are numerical results. The lines show the analytical results: the velocity $V$ [Eq.~(\ref{eq:V})] and the angular velocity $\Omega$ [Eq.~(\ref{eq:Omega0})] within the linear response. Note the sign change of the velocity across $\delta_s = 0$ and the maximum of the angular velocity at $\delta_s = 0$.}
\label{fig:fig3}
\end{figure}

Figure~\ref{fig:fig2}(a) and (b) show the domain-wall velocity $V$ and the angular velocity $\Omega$ as a function of the current density $J \le 1 \times 10^{12}$ A/m$^2$ for various values of the net spin density $\delta_s$ shown in Table~\ref{tab:tab1}. The analytical results shown as lines and the numerical results shown as symbols agree well for the current densities $J \lesssim 5 \times 10^{11}$ A/m$^2$. Figure~\ref{fig:fig3}(a) and (b) show the velocity and the angular velocity as a function of the net spin density $\delta_s$ for various current densities. Two main features predicted by the theory, the sign change of the domain-wall velocity across $\delta_s = 0$ and the maximum angular velocity $\Omega$ at $\delta_s = 0$, are demonstrated in the numerical results.

\section{STT-induced domain-wall dynamics at $T_A$}
\label{sec:sec3}

In this section, we study the STT-induced dynamics of a domain wall exactly at $T_A$, where the net spin density vanishes and thus the nature of the magnetic dynamics is purely antiferromagnetic, by applying a charge-current density up to $4 \times 10^{12}$ A/m$^2$ to look for novel phenomena.

\begin{figure}
\includegraphics[width=\columnwidth]{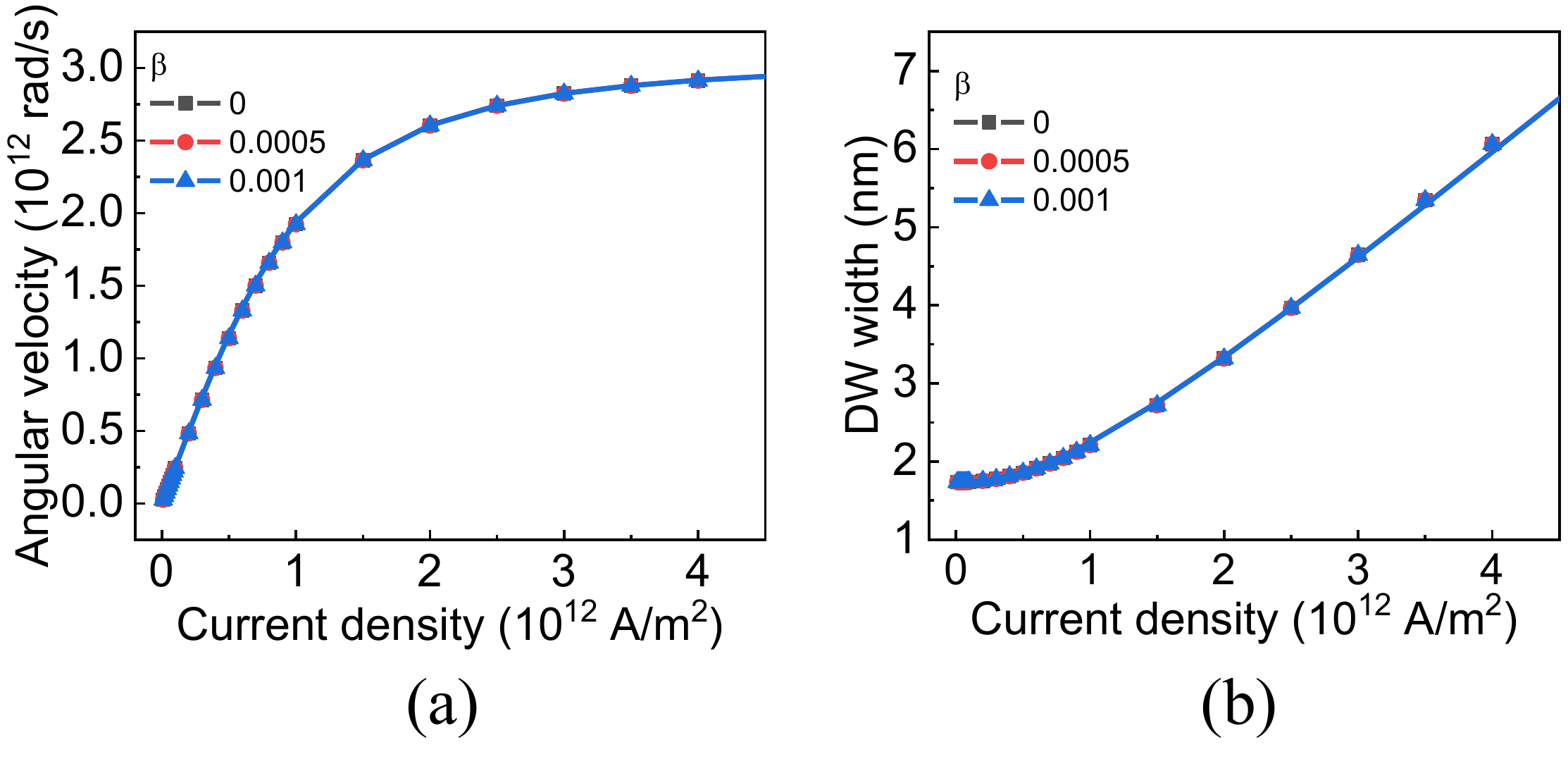}
\caption{(a) The domain-wall angular velocity and (b) the domain-wall width as functions of the current density for three different values of the nonadiabatic STT parameter $\beta = 0, 0.0005, 0.001$ at the angular momentum compensation point $\delta_s = 0$. The symbols are numerical results. The solid lines show the analytical results: the angular velocity $\Omega$ [Eq.~(\ref{eq:Omega})] and the width $\lambda$ [Eq.~(\ref{eq:lambda})].}
\label{fig:fig4}
\end{figure}

\subsection{Angular velocity of a domain wall}

Let us first discuss the numerical results on the domain-wall angular velocity from atomistic spin simulations performed with three different values of the nonadiabatic STT parameter $\beta = 0, 0.0005$, and $0.001$. Figure~\ref{fig:fig4}(a) shows the angular velocity $\dot{\Phi}$ as a function of the current density. The angular velocity increases linearly as the current increases for the small current density as predicted by Eq.~(\ref{eq:Omega0}), but it deviates from the equation for high current densities by showing the saturation.

This observed saturation of the angular velocity can be understood as the effect of the change of the domain-wall width as follows. The width of the static domain wall is given by $\lambda_0 = \sqrt{A/K}$ determined by the competition between the exchange energy $\propto A$ and the easy-axis anisotropy $\propto K$. When the domain wall is precessing uniformly at the angular velocity $\Omega$ in the laboratory frame, the effective easy-axis anisotropy in the spin frame rotating at the domain-wall angular velocity $\Omega$ is given by $K_\text{eff} = K - \rho \Omega^2$ as shown in Ref.~\cite{KimPRB2014}: the uniform spin rotation about the $z$ axis in the laboratory frame gives rise to the effective magnetic field along the $z$ axis in the rotating spin frame of reference (which is analogous to the centrifugal force in a rotating frame of reference), which in turn decreases the easy-axis anisotropy as known for magnets with antiferromagnetic coupling~\cite{BaltzRMP2018}. Therefore, the width of the domain wall rotating at the angular velocity $\Omega$ is given by
\begin{equation}
\label{eq:lambda}
\lambda = \frac{\lambda_0}{\sqrt{1 - (\Omega / \omega_0)^2}} \, ,
\end{equation}
where $\omega_0 \equiv \sqrt{K / \rho}$ is the spin-wave gap at $T_A$~\cite{OhPRB2017}. By solving Eq.~(\ref{eq:Omega0}) with $\lambda_0$ replaced by $\lambda$ [Eq.~(\ref{eq:lambda})] for $\Omega$, we obtain the domain-wall precession frequency as a function of the current density:
\begin{equation}
\Omega = \frac{P J / \alpha \lambda_0 s}{\sqrt{1 + (P J / \alpha \lambda_0 s \omega_0)^2}} \, , \label{eq:Omega}
\end{equation}
and the domain-wall width:
\begin{equation}
\lambda = \lambda_0 \sqrt{1 + (P J / \alpha \lambda_0 s \omega_0)^2} \, . \label{eq:lambda}
\end{equation}
The obtained expression for the angular velocity $\Omega$ [Eq.~(\ref{eq:Omega})] is reduced to the linear-response result [Eq.~(\ref{eq:Omega0})] when the quadratic effects in the current density $J$ is neglected. Note that the angular velocity converges to the spin-wave gap $\omega_0 \approx 3 \times 10^{12}$ rad/s as the current density increases, but can never exceed it. The sold lines in Figs.~\ref{fig:fig4}(a) and (b) show the analytical solutions for the angular velocity $\Omega$ [Eq.~(\ref{eq:Omega})] and width $\lambda$ [Eq.~(\ref{eq:lambda})], respectively for several values of $\beta$. They agree well with the simulation results shown as the symbols. 

\begin{figure}
\includegraphics[width=\columnwidth]{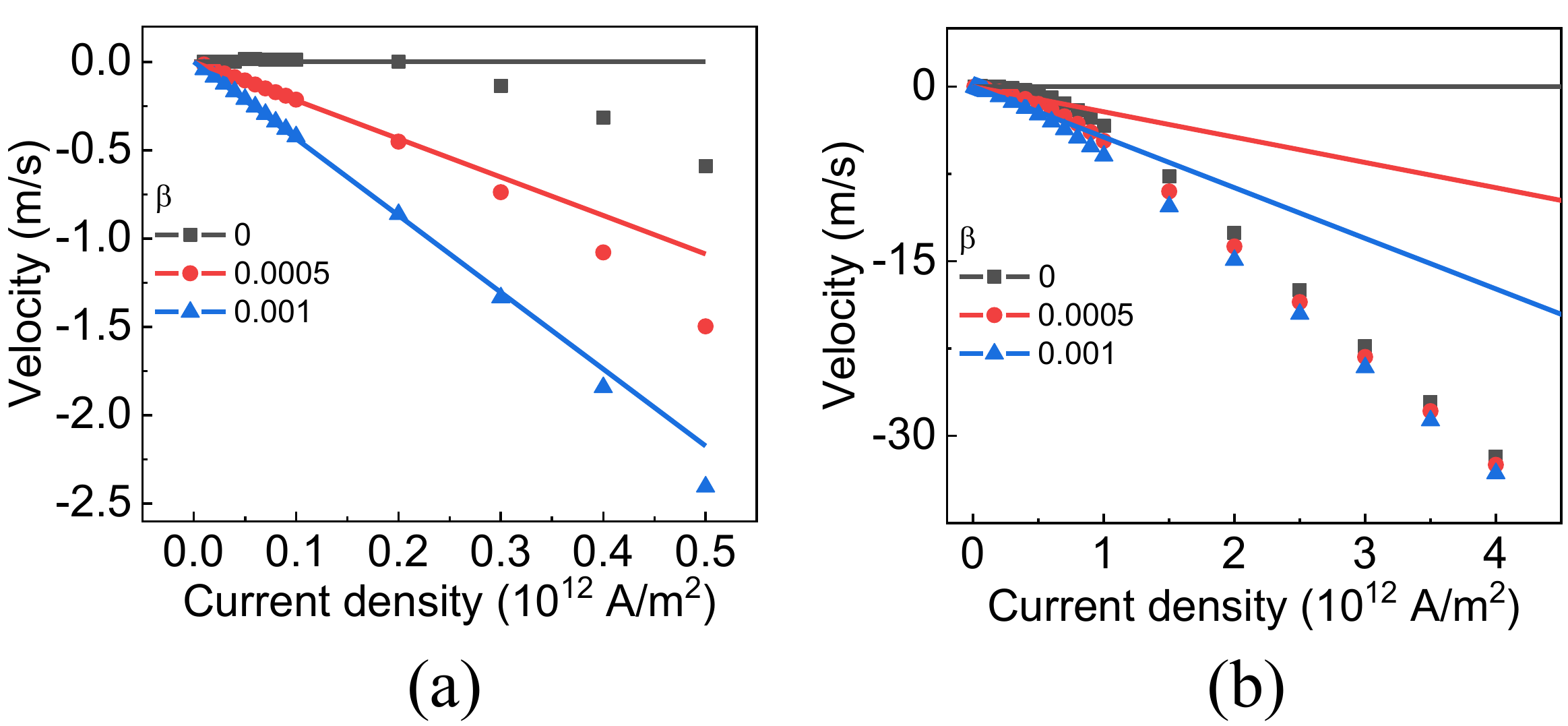}
\caption{(a) and (b) The domain-wall velocity as a function of the current density for three different values of the nonadiabatic STT parameter $\beta = 0, 0.0005, 0.001$ at the angular momentum compensation point $\delta_s = 0$. The symbols are simulation results. The sold lines show the analytical results for the domain-wall velocity $V$ within the linear response [Eq.~(\ref{eq:V})].}
\label{fig:fig5}
\end{figure}

\subsection{Velocity of a domain wall}

Let us now turn to the STT-induced translation motion of a domain wall at large currents. Figure~\ref{fig:fig5}(a) shows the domain-wall velocity $V$ as a function of the current density up to $0.5 \times 10^{12}$ A/m$^2$. For relatively small currents $J \leq 0.2 \times 10^{12}$ A/m$^2$, the simulation results shown as symbols are well explained by the linear-response analytical result $V = - \beta P J / \alpha s$ [Eq.~(\ref{eq:V})] shown as solid lines. However, Fig.~\ref{fig:fig5}(b), where the current density as large as $4 \times 10^{12}$ A/m$^2$ is applied, shows that the domain-wall velocity starts to deviate significantly from Eq.~(\ref{eq:V}) for the current density $J \gtrsim 1 \times 10^{12}$ A/m$^2$. This deviation is not due to the current-induced change of the domain-wall width since $V = - \beta P J / \alpha s$ does not depend on $\lambda$. There are two notable features. First, even when the nonadiabatic torque is absent $\beta = 0$, the domain wall exhibits a finite velocity at high current densities, which disagrees with the known results for antiferromagnetic domain-wall motion obtained within the linear response~\cite{HalsPRL2011}. Secondly, as the current density increases, the domain-wall velocities corresponding to three different values of $\beta$ appear to converge on the one universal line, suggesting that it is not the nonadiabatic STT $\propto \beta P J$ but the adiabatic STT $\propto PJ$ that plays a main role in the observed domain-wall velocity at high current densities. Our numerical result demonstrates a limitation of the linear-response theory for the STT-induced domain-wall motion at high currents. We leave a theoretical understanding of the observed domain-wall velocity at higher currents as a future research topic.

\section{Discussion}
\label{sec:sec4}

To sum up, we have studied the STT-induced dynamics of a domain wall in ferrimagnets theoretically and numerically. The domain-wall velocity changes it sign across $T_A$ due to the sign change of the net spin density, giving rise to a phenomenon unique to ferrimagnets that cannot be found in ferromagnets and antiferromagnets. The angular velocity of a domain wall is shown to exhibit its maximum at $T_A$, which can be understood as the effect of the STT-induced accumulation of the nonequilibrium spin density inside the domain wall. At higher currents, we have found numerically that the domain-wall velocity can significantly deviate from the linear-response result, calling for the development of a more general theory for the dynamics of a domain wall subjected to strong currents.

In this work, we have focused on the effects of STT on the dynamics of a domain wall in ferrimagnets. The reciprocal effects of a spin texture on a current are known to give rise to intriguing phenomena in ferromagnets such as the generation of electromotive force by domain-wall precession~\cite{BergerPRB1986, VolovikJPC1987, YangPRL2009, YangPRB2010} and the topological Hall effect in skyrmion crystal phases~\cite{BinzPB2008, LeePRL2009, NeubauerPRL2009}. The corresponding effects in ferrimagnets would be worth being investigated in the future. In addition, our understanding of STT in antiferromagnetically-coupled magnetic systems can be advanced further by pursuing the microscopic theory for the spin-charge interaction in ferrimagnets as has been done for ferromagnets within the Stoner model or the s-d model for itinerant ferromagnetism~\cite{TserkovnyakJMMM2008}. More generally, we envision that the research on the spin dynamics as well as the spin-charge interaction in ferrimagnets will lead us to more unified understanding of magnetic phenomena spanning various types of magnetic materials including ferromagnets and antiferromagnets as two special cases, and also it will facilitate the advancement of ferrimagnetic spintronics aiming at easily-controllable high-speed devices by combining the advantages of ferromagnetic and antiferromagnetic devices.

\begin{acknowledgements}
D.H.K was supported by the National Research Council of Science \& Technology (NST) Research Fellowship for Young Scientist of the National Research Council of Science \& Technology (NST), the POSCO Science Fellowship of POSCO TJ Park Foundation, the Korea Institute of Science and Technology (KIST) institutional program (No. 2E29410 and 2E30600), and the National Research Council of Science \& Technology (NST) grant (No. CAP-16-01-KIST) funded by the Korea government (Ministry of Science and ICT). K.J.K. was supported by the National Research Foundation of Korea (NRF) grant funded by the Korean government (MSIP) (Grant No. NRF-2016R1A5A1008184). K.W.M. and S.Y. were supported by the National Research Foundation of Korea (NRF-2019M3F3A1A02072478), the National Research Council of Science \& Technology (NST) grant (No. CAP-16-01-KIST) by the Korea government (MSIP), the Future Materials Discovery Program through the National Research Foundation of Korea (No. 2015M3D1A1070467). K.J.L. was supported by the KIST Institutional Program (No. 2V05750). S.K.K. was supported by Young Investigator Grant (YIG) from Korean-American Scientists and Engineers Association (KSEA) and Research Council Grant URC-19-090 of the University of Missouri.
\end{acknowledgements}

\bibliography{/Users/kimsek/Dropbox/School/Research/master}

\end{document}